\documentclass[conference]{IEEEtran}
\usepackage{cite}
\usepackage{amsmath,amssymb,amsfonts}
\usepackage{algorithmic}
\usepackage{graphicx}
\usepackage{textcomp}
\usepackage{xcolor}
\def\BibTeX{{\rm B\kern-.05em{\sc i\kern-.025em b}\kern-.08em
    T\kern-.1667em\lower.7ex\hbox{E}\kern-.125emX}}

\usepackage{booktabs}
\usepackage{array}

\usepackage{makecell}

\usepackage{subcaption}

\begin{document}

\title{Circuit-Based Dispersion Analysis of Periodic Cross-Shaped Unit Cells with Dirac Characteristics}

\author{\IEEEauthorblockN{Mahyar Mehri Pashaki}
\IEEEauthorblockA{\textit{Department of Electrical Engineering} \\
\textit{Sharif University of Technology}\\
Tehran, Iran\\
mahyar.mehri.p@gmail.com}
\and
\IEEEauthorblockN{Mohammad Hossein Koohi Ghamsari}
\IEEEauthorblockA{\textit{Department of Electrical Engineering} \\
	\textit{Sharif University of Technology}\\
	Tehran, Iran\\
	mohammadghamsari18@gmail.com}
\and
\IEEEauthorblockN{Mohammad Memarian}
\IEEEauthorblockA{\textit{Department of Electrical Engineering} \\
	\textit{Sharif University of Technology}\\
	Tehran, Iran\\
	mmemarian@sharif.edu}

}

\maketitle

\begin{abstract}
In this paper, the dispersion behavior of periodic cross-shaped unit cells is investigated with emphasis on Dirac-type dispersion using an equivalent circuit modeling approach. Two unit-cell geometries, including a simple cross structure and a modified configuration with a central patch, are analyzed using full-wave eigenmode simulations and scattering-parameter-based dispersion extraction. An equivalent circuit model is developed to capture the dominant coupling mechanisms and accurately reproduce the dispersion characteristics near the $\Gamma$-point. The results demonstrate that the proposed circuit model exhibits good agreement with full-wave simulations and provides a physically intuitive framework for analyzing stopband behavior and mode degeneracy. Furthermore, the limitations of S-parameter-based dispersion extraction in the presence of low-Q radiative modes are highlighted.

\end{abstract}


\section{Introduction}

The development of automotive radar sensors and antennas, particularly for autonomous driving systems operating in the 77GHz band, poses significant design and implementation challenges. Directional antennas are typically employed in automotive radar applications, with phased-array antennas being conventional candidates due to their beam-steering capability. However, the complexity and cost associated with designing the feeding network for large arrays motivate alternative solutions. Among these, leaky-wave antennas (LWAs) offer a simpler and more effective approach to meet the requirements of automotive radar systems \cite{mesa2020simulation, giusti2022efficient}. LWAs can radiate directive beams through frequency scanning using a single-point feed, which simplifies fabrication and substantially reduces feed-network complexity compared with phased-array structures \cite{dorrah2018two,dorrah2016pencil}. 

Compared to conventional phased arrays, LWAs are particularly attractive for millimeter-wave automotive radar due to their compact footprint, low-profile implementation, and reduced sensitivity to feeding-network losses at high frequencies \cite{dorrah2016modal,dorrah2018high}. These characteristics make LWAs well suited for cost-sensitive and space-constrained radar platforms, where ease of integration and fabrication robustness are critical design considerations. As a result, LWAs have emerged as a promising alternative architecture for next-generation automotive radar antennas \cite{memarian2015dirac,rezaee2020analytical}.

In this work, Dirac-type leaky-wave antennas (DLWAs) are considered as promising candidates for automotive radar applications. These antennas feature a simple geometry, low design complexity, and cost-effective fabrication \cite{escobar2023homogenization}. Although several DLWA designs have been studied at lower frequency bands, their extension to automotive radar frequencies around 77~GHz has not been explored yet. The proposed approach focuses on implementing such Dirac-type leaky-wave structures on planar microstrip platforms to ensure ease of integration with commercial radar sensor modules \cite{rahmeier2021complex}. For accurate physical insight, precise circuit modeling of the unit cells is required \cite{garcia2025multimodal}. Circuit-based representations provide a quantitative understanding of the dispersion behavior of periodic structures and allow prediction of bandgap opening or closure, which is critical for beam radiation near broadside \cite{friedman1990principles, dudley2015mathematical}.

Dispersion engineering plays a central role in the design of periodic leaky-wave structures, as the frequency–wavenumber relationship directly governs beam direction, scanning behavior, and the existence of stopbands \cite{otto2014transversal, dyab2015interpretation, otto2012complex}. In DLWAs, accurate dispersion control at the unit-cell level is essential to enforce linear dispersion characteristics around the Dirac point and to avoid the formation of a broadside stopband. Consequently, reliable dispersion modeling becomes a key requirement for the successful design of such antennas.

Furthermore, dispersion extraction plays an essential role in evaluating periodic leaky-wave structures. Several studies have addressed extracting the dispersion diagrams from scattering parameters, each method having its own advantages and limitations. Understanding this behavior enables reliable prediction of radiating modes and bandgap properties, and provides useful insight into the underlying physical mechanisms governing wave propagation in periodic antenna structures.

Finally, this paper introduces a new unit cell exhibiting Dirac-type dispersion with the capability of achieving a closed bandgap. Such a configuration enables broadside radiation, making it ideally suited for automotive radar applications. In addition, a highly accurate circuit model is developed to characterize the dispersion behavior of the proposed periodic structure, showing excellent agreement with full-wave electromagnetic simulations.

\section{Unit-Cell Circuit Model}

\subsection{Unit-Cell Structure}

In this section, we introduce a modified cross-shaped unit cell incorporating a central square patch, designed to exhibit Dirac-type dispersion characteristics. This geometry serves as the reference model for evaluating dispersion-extraction procedures and is particularly suitable for enabling broadside radiation without stopband formation. The cell is implemented on a 10‑mil RO4003C substrate with a dielectric loss tangent of 0.0027, and its dimensions are chosen to be approximately half of the guided wavelength at the 77~GHz operating frequency. This frequency band is widely adopted in automotive radar systems for advanced driver-assistance and autonomous driving applications, making the proposed unit cell highly relevant for practical integration in planar microstrip antenna platforms. Under the excitation and boundary conditions illustrated in Fig.~1, the unit cell supports the accidental degeneracy of modes at the $\Gamma$ point, resulting in a closed bandgap and a dispersion profile characteristic of Dirac cones, which is advantageous for continuous beam scanning in periodic leaky-wave antenna designs.

\begin{figure}[!t]
	\centering
	
	\includegraphics[scale=0.27,trim=5.5cm 19cm 2cm 1cm,clip=true]{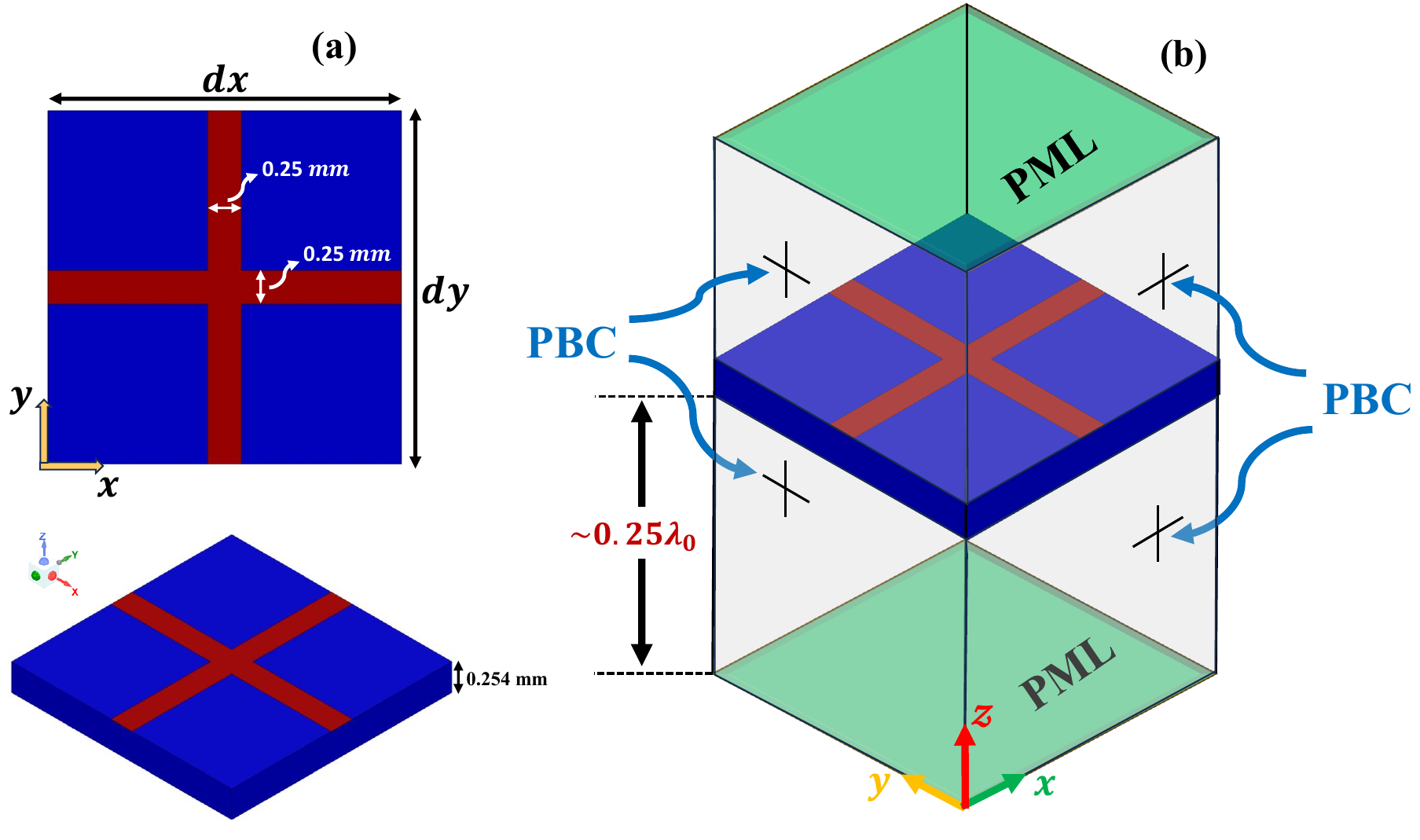}
	
	\captionsetup{
		font=footnotesize, 
		singlelinecheck=false, 
		margin={0pt,0pt} 
	}	
	\caption{(a) Structure of the simple cross-line unit-cell. Dimensions of the square unit-cell are selected as dx=dy= $\lambda/2$. (b) The simulation setup of the unit-cell. Parameter $\lambda_0$ is the free space wavelength calculated as $\lambda_0 = c/f$, where c and f are $3\times10^8$ and $f=77 GHz$, respectively.}
	\label{First_Unit_Cell}
\end{figure}

As discussed earlier, two distinct approaches are adopted to extract the dispersion characteristics of the periodic structure. The first method relies on a full-wave eigenmode analysis, in which the unit cell is simulated in Ansys HFSS under periodic boundary conditions, and the Bloch propagation constant is obtained directly from the computed eigenfrequencies and field distributions. The simulation setup used for this analysis is illustrated in Fig.~1(b).

In the second approach, a driven analysis is performed with the primary objective of validating the circuit-based model against full-wave simulations. To this end, the periodic unit cell is modeled in HFSS as a four-port network, where the transverse directions are terminated using additional ports. This driven setup allows the scattering parameters of the unit cell to be computed and directly compared, in both magnitude and phase, with those predicted by the proposed equivalent circuit model. Once a consistent agreement is established at the scattering-parameter level, the extracted S-parameters are post-processed and reduced to an equivalent two-port representation. The Bloch wavenumber is then obtained from this reduced network, enabling a final comparison between the dispersion diagrams derived from the circuit-based model and those obtained using the eigenmode analysis.

\begin{figure}[!t]
	\centering
	
	\includegraphics[scale=0.4,trim=6cm 2cm 6cm 1.5cm,clip=true]{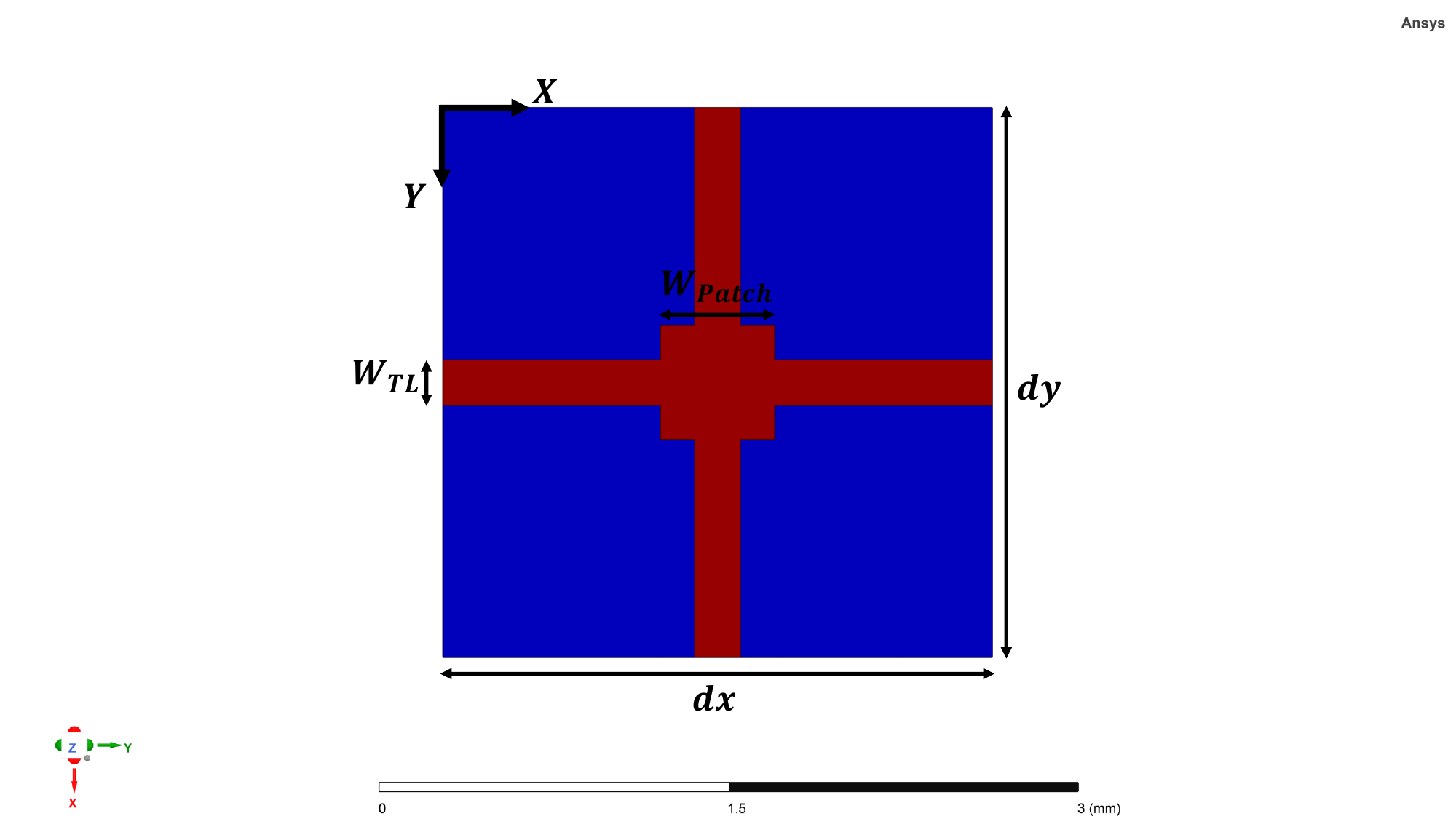}
	
	\captionsetup{
		font=footnotesize, 
		singlelinecheck=false, 
		margin={0pt,0pt} 
	}	
	\caption{Geometry of the cross-shaped unit cell with a central patch. The overall cell size is approximately half of the operating wavelength.}
	\label{Second_Unit_Cell}
\end{figure}

The second unit-cell configuration, shown in Fig.~2, incorporates a central patch at the intersection of the cross arms. This structural modification increases the inter-arm coupling and enables tuning of the dispersion characteristics toward conditions associated with Dirac-like behavior. Such a configuration allows the periodic cell to operate near a transition point between open and closed stopbands, providing a useful platform for investigating the evolution of the dispersion diagram. In this study, the cross-with-patch unit cell is used to evaluate how the equivalent circuit model can capture the transition between these two regimes. By adjusting the size of the central patch, the modal characteristics of the structure particularly the position of the frequency bands associated with the Dirac point  can be systematically shifted, enabling a direct comparison between the dispersion predicted by the circuit model and that obtained from the eigenmode analysis.

\subsection{CIRCUIT MODEL Design}
For the modeling and design of the different types of Dirac Leaky Wave(DLW) unit-cells, the analytical methods do not always provide an accurate response to the dispersion behavior of DLW structures. Two other alternatives are full-wave simulations or circuit models. Full-wave simulations are time-consuming and complex, especially for analyzing modern high-frequency mm-wave structures. On the other hand, circuit models have offered a simple and intuitive approach to the design and analyses of periodic structures unit-cells. 

In this section, we propose accurate circuit models for the
proposed unit cells to efficiently design customized DLW unit
cells at the desired frequency. These circuit models accurately
predict the dispersion behavior of the DLW unit cells. Furthermore, these models allow us to easily adjust the cell design
parameters and perform appropriate optimization procedures,
paving the way for an efficient design and realization of the
DLW structures.

Fig.~3 illustrates the equivalent circuit model of the simple cross-line DLW unit cell shown in Fig.~1. In this model, the microstrip cross patch represents the central intersection of the unit cell, while four microstrip lines are used to model the metallic arms. The circuit is excited through four matched 50‑$\Omega$ ports. In addition, the equivalent circuit of the microstrip cross‑patch component, composed of lossless inductors and capacitors, is shown in Fig.~3(b). A detailed derivation of the circuit parameters associated with the cross structure can be found in \cite{gupta1981computer}.

\begin{figure}[!t]
	\centering
	
	\includegraphics[scale=0.35,trim=5cm 5cm 5cm 4.2cm,clip=true]{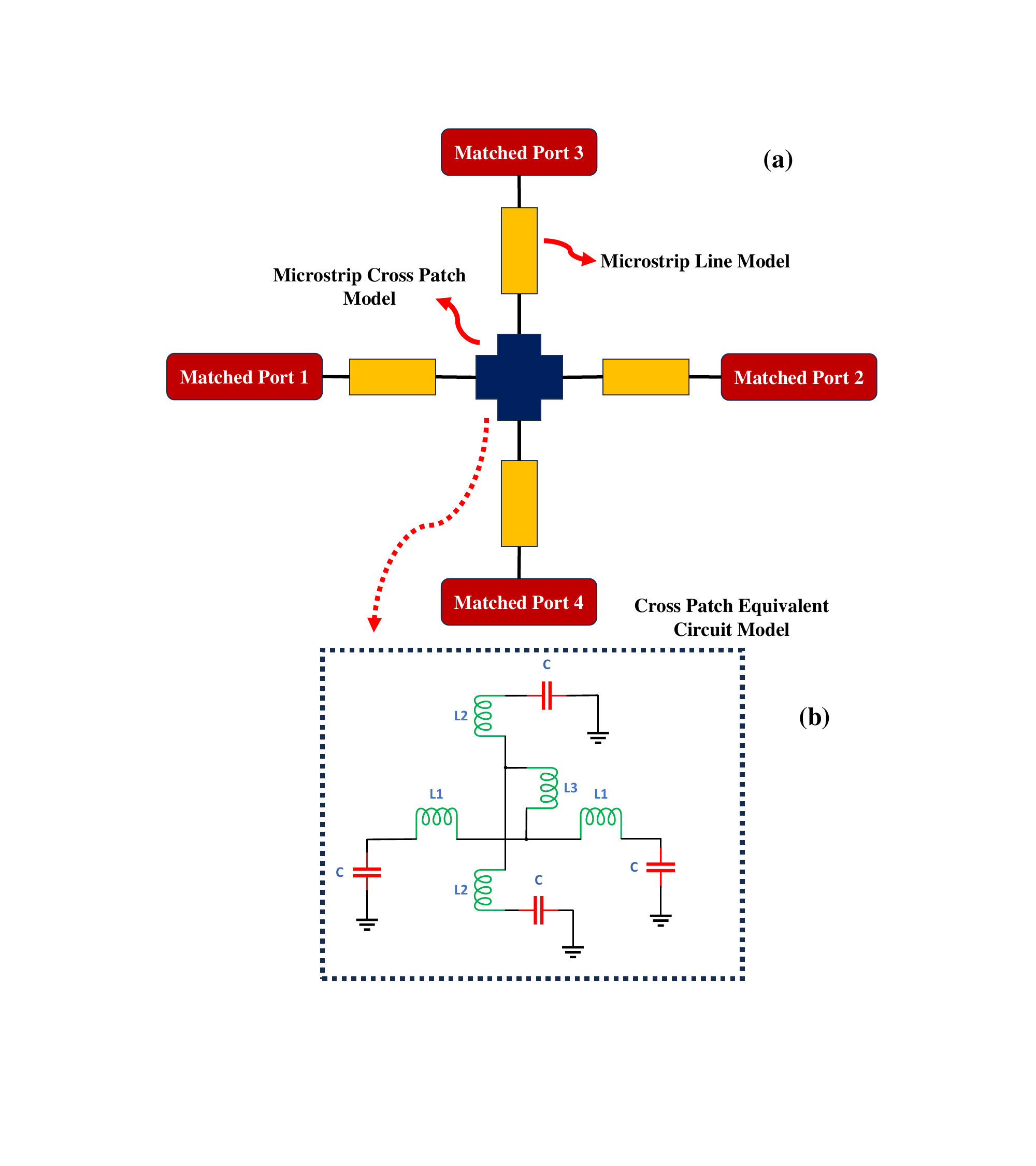}
	
	\captionsetup{
		font=footnotesize, 
		singlelinecheck=false, 
		margin={0pt,0pt} 
	}	
	\caption{Proposed circuit model for the simple cross-line DLW unit-cell of Fig. 2. (a) Schematic Representation. (b) Equivalent circuit of the microstrip cross patch component.}
	\label{model_simple_crossline}
\end{figure}

\section{Results and Dispersion Validation}
To verify the accuracy of the proposed circuit model, we followed the following steps:

1-	First, the scattering parameters amplitude and phase of the simple cross-line LWD unit-cell of Fig. \ref{First_Unit_Cell} are calculated using the circuit model of Fig. \ref{model_simple_crossline}. In addition, the Modal solver of ANSYS HFSS 2022 is utilized to calculate the scattering parameters of the unit-cell and the results are compared.

2-	The dispersion curve of the unit-cell is calculated using the scattering parameters obtained from the circuit model in the previous step. Also, the dispersion diagram of the unit-cell is extracted using the Eigenmode Solver of HFSS. The results of dispersion curves obtained from both methods are compared in Fig. \ref{Fig20}.

To validate the proposed circuit formulation, we first investigate the scattering-parameter response of the simple cross-line unit cell shown in Fig.~1 using the equivalent circuit model illustrated in Fig.~3. The magnitude and phase of the scattering parameters obtained from the circuit model are initially compared with those extracted from full-wave simulations. This preliminary comparison shows a reasonable level of agreement, indicating that the proposed circuit topology captures the dominant electromagnetic behavior of the unit cell. However, to achieve a more accurate correspondence with the full-wave results, four additional capacitive elements are introduced along the arms of the circuit model in Fig.~3. By appropriately tuning the values of these capacitors, the circuit model is able to closely reproduce the scattering-parameter response of the simple cross structure, resulting in a refined model that exhibits excellent agreement in both magnitude and phase.

Next, the same circuit model shown in Fig.~3 is applied to the cross unit cell with a central patch illustrated in Fig.~2. The scattering parameters of this configuration, obtained from the circuit model and full-wave simulation, are compared in Fig.~5(c) and Fig.~5(d) in terms of magnitude and phase, respectively. As can be clearly observed, the agreement between the circuit-based results and the full-wave simulations is significantly degraded for this unit cell. This noticeable mismatch indicates that the original circuit model of Fig.~3 is no longer sufficient to accurately capture the scattering-parameter behavior of the cross-with-patch configuration.

\begin{figure}[!t]
	\centering
	
	\includegraphics[scale=0.19,trim=5cm 3cm 5cm 3cm,clip=true]{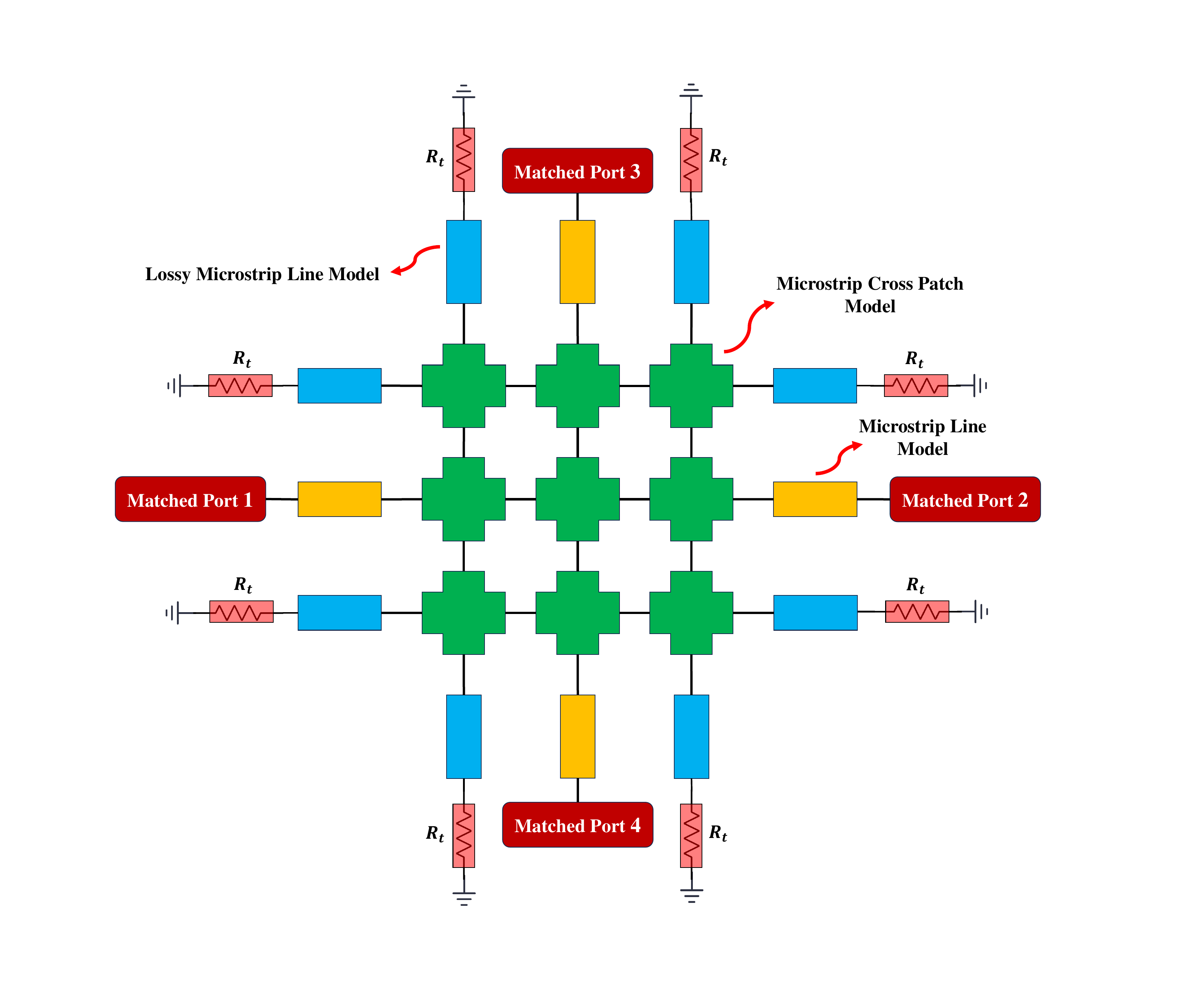}
	
	\captionsetup{
		font=footnotesize, 
		singlelinecheck=false, 
		margin={0pt,0pt} 
	}	
	\caption{The modified proposed circuit model for the cross-line DLW unit-cell with a large patch that satisfies the radiation condition.}
	\label{second_patch_splitter}
\end{figure}

\begin{figure}[!t]
	\centering
	
	\includegraphics[width=\columnwidth,trim=3cm 0cm 3cm 0cm,clip=true]{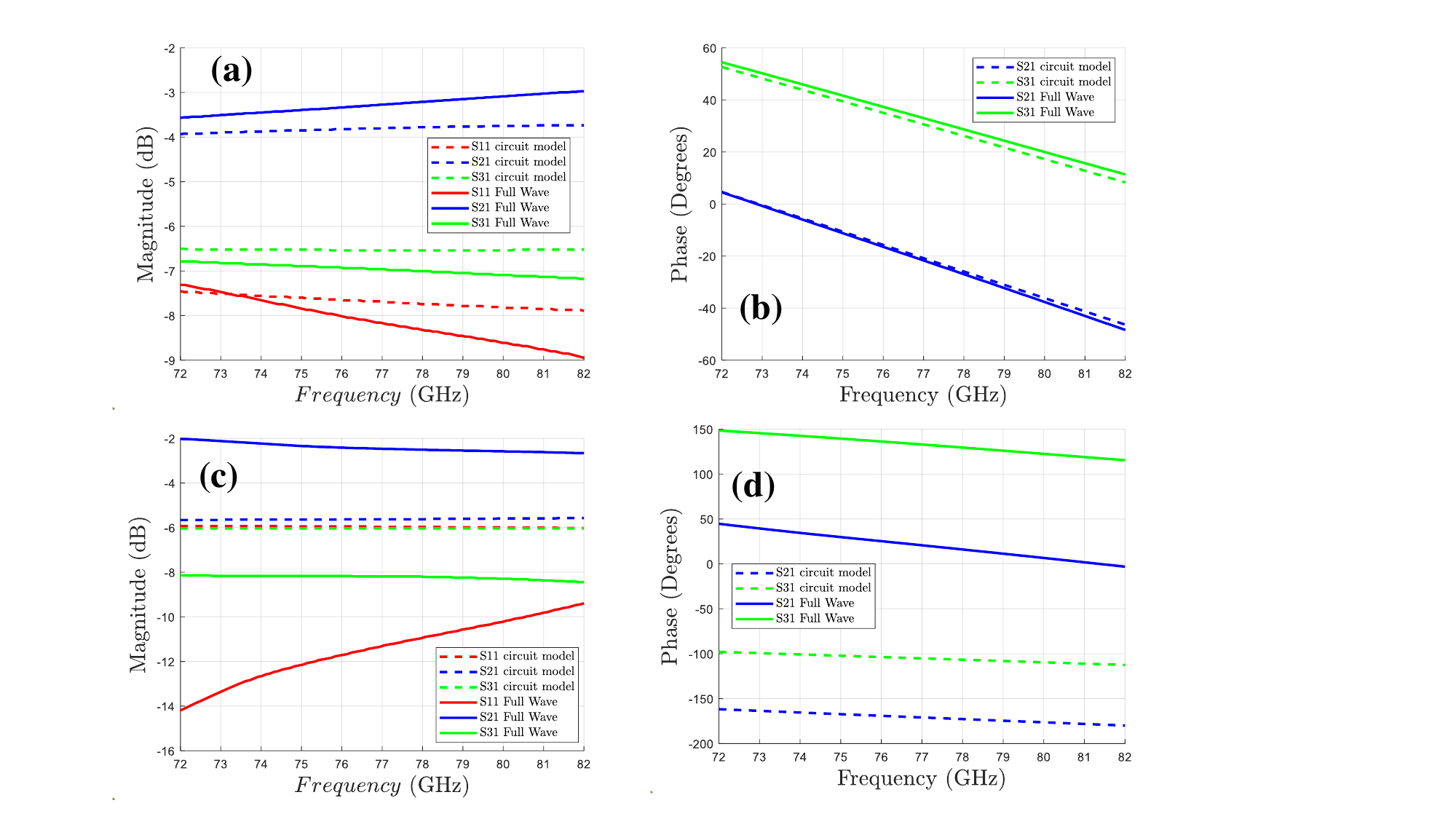}
	
	\captionsetup{
		font=footnotesize, 
		singlelinecheck=false, 
		margin={0pt,0pt} 
	}	
	\caption{Comparison of the scattering parameters obtained from the equivalent circuit model and full-wave simulations. (a),(b) Magnitude and phase comparison for the unit cell structure shown in Fig.~1, modeled using the equivalent circuit of Fig.~3. (c),(d) Magnitude and phase comparison for the unit cell structure shown in Fig.~2, analyzed using the same equivalent circuit model of Fig.~3.
		.}
	\label{CircuitModel_SParam}
\end{figure}

Since even after tuning the parasitic capacitors the circuit model presented in Fig.~3 did not provide a sufficiently accurate match with the full-wave results, a modified approach is introduced. In this method, the central patch section of the unit cell is divided into multiple identical sub‑patches, as illustrated in Fig.~4. This segmentation technique effectively improves the modeling accuracy by satisfying the geometrical constraints associated with the patch width and height. In addition, parasitic capacitors are still incorporated between the junctions of these sub‑patches to fine‑tune the overall response. The complete configuration of the improved circuit model is shown in Fig.~4, demonstrating the refined representation of the central patch and its coupling network.

Using this improved modeling approach, the scattering-parameter behavior of the combined patch-and-cross structure is presented in Fig.~6. As observed in this figure, the circuit model accurately follows both the magnitude and phase responses obtained from the full-wave simulation. This close agreement confirms that the refined model provides a reliable representation of the electromagnetic behavior of the unit cell over the frequency range of interest.

\begin{figure}[!t]
	\centering
	
	\includegraphics[width=\columnwidth,trim=5.5cm 9.5cm 3cm 0cm,clip=true]{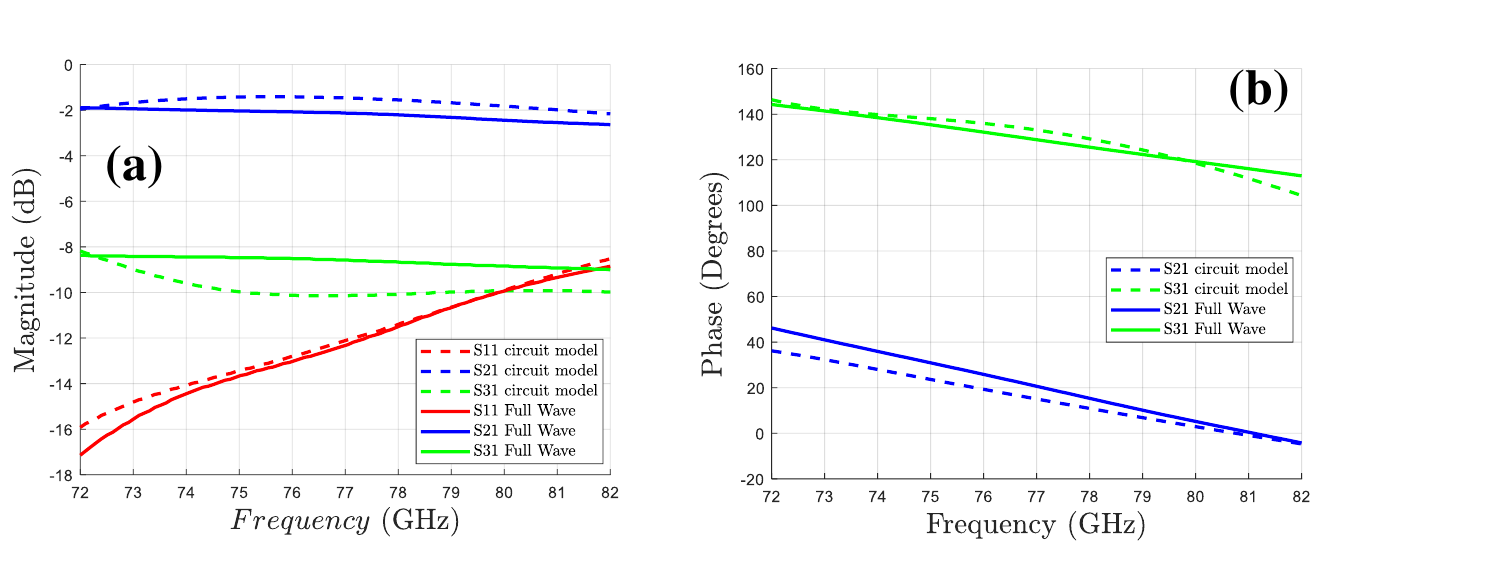}
	
	\captionsetup{
		font=footnotesize, 
		singlelinecheck=false, 
		margin={0pt,0pt} 
	}	
	\caption{Comparison of the scattering parameters obtained from the equivalent circuit model and the full-wave simulation for the unit cell structure shown in Fig.~2, using the improved circuit model presented in Fig.~4.}
	\label{CircuitModel_Dispersion}
\end{figure}

Finally, the dispersion diagrams obtained from the full-wave simulation and the proposed circuit model are illustrated in Fig.~7. The results are compared for two cases: the open-bandgap and closed-bandgap conditions. As can be observed, the two dispersion curves show very good agreement in both cases. This strong consistency indicates that the developed circuit model accurately follows the full-wave behavior and validates the correctness and effectiveness of the proposed modeling approach. Furthermore, the three-dimensional radiation pattern of the final antenna design, which is suitable for automotive radar applications, is presented in Fig.~8.

\begin{figure}[!t]
	\centering
	
	\includegraphics[scale=0.3,trim=1cm 4cm 4cm 2cm,clip=true]{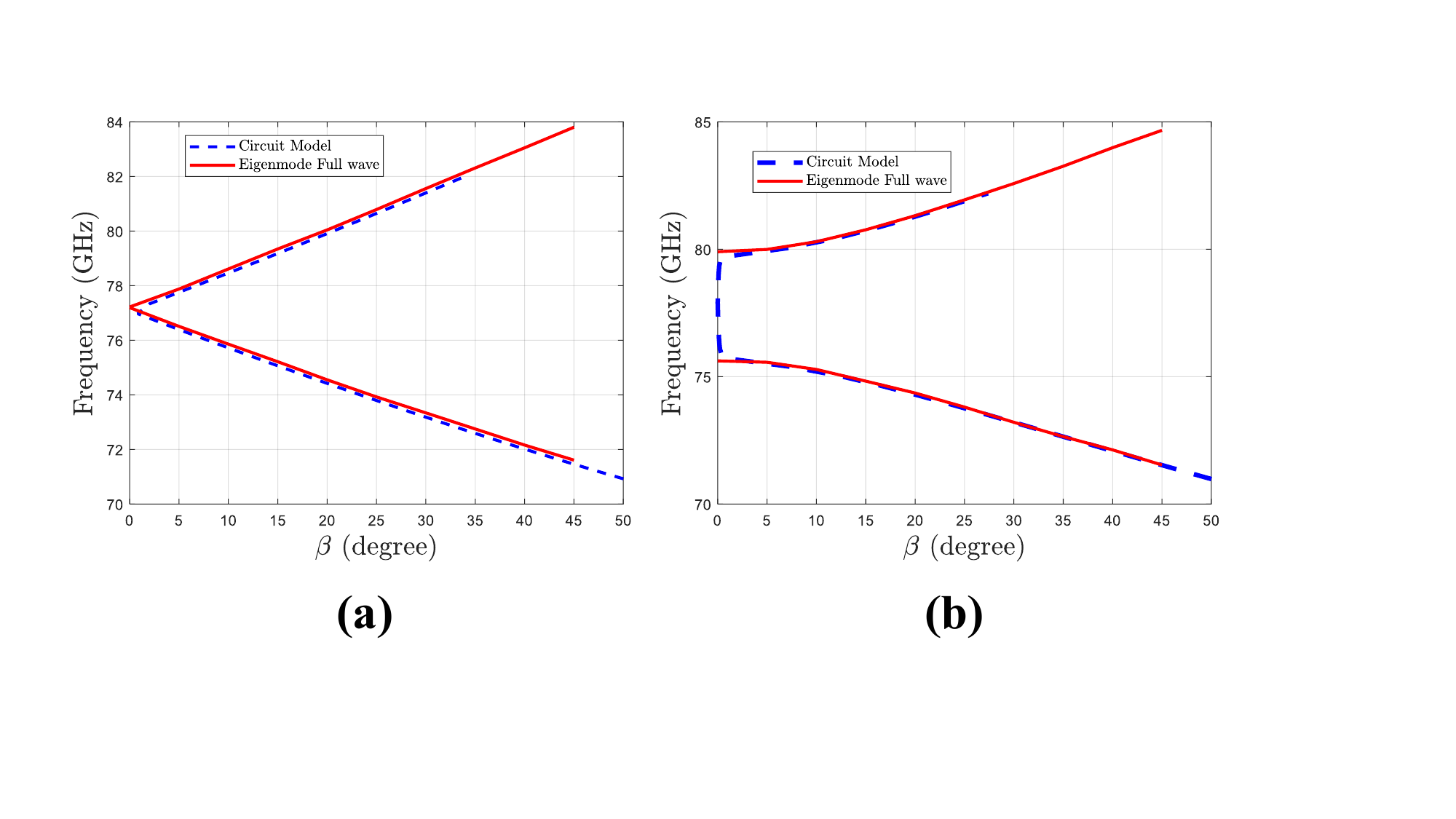}
	
	\captionsetup{
		font=footnotesize, 
		singlelinecheck=false, 
		margin={0pt,0pt} 
	}	
	\caption{Comparison of the dispersion diagrams obtained from the equivalent circuit model and the full-wave eigenmode analysis for two unit-cell structures exhibiting open-bandgap and closed-bandgap conditions.}
	\label{Fig20}
\end{figure}

\begin{figure}[!t]
	\centering
	
	\includegraphics[scale=0.35,trim=6cm 4cm 6cm 2cm,clip=true]{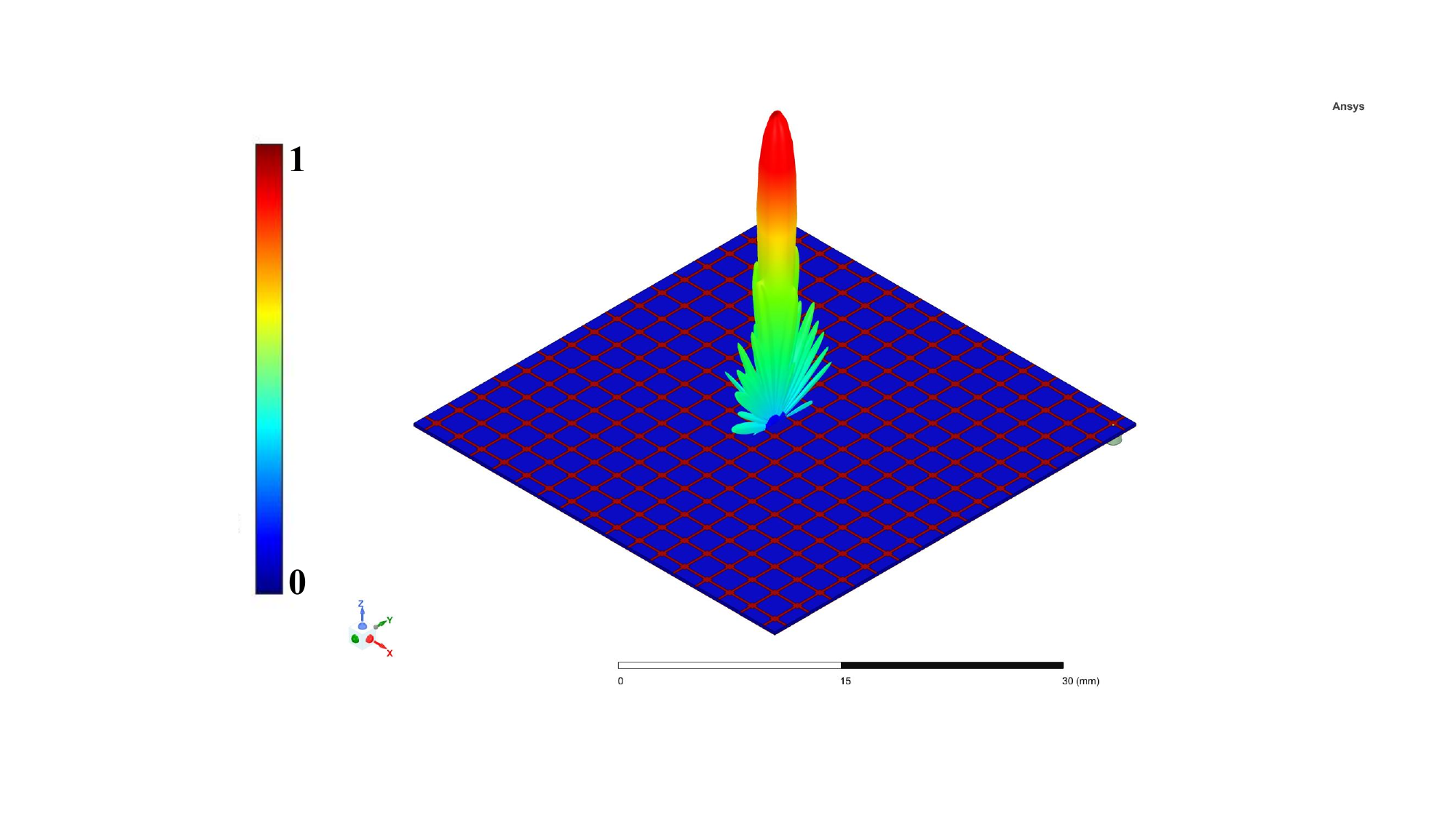}
	
	\captionsetup{
		font=footnotesize, 
		singlelinecheck=false, 
		margin={0pt,0pt} 
	}	
	\caption{Three-dimensional radiation pattern of the final antenna at 77GHz,
		demonstrating a highly directive broadside beam suitable for automotive radar
		applications.}
	\label{Pattern3D}
\end{figure}

Finally, it should be emphasized that these dispersion-extraction approaches may exhibit different behaviors depending on the electromagnetic characteristics of the considered unit-cell structures. In many recent studies, various techniques are introduced to extract dispersion diagrams from the scattering parameters of a single unit cell and are directly compared with full-wave driven-modal results. However, under certain conditions, the presence of radiative modes in the structure can cause a significant portion of the modal energy to leak out through radiation rather than being fully captured by the ports. This effect may lead to a misleading interpretation of the dispersion behavior, where, for example, a unit cell exhibiting an open-bandgap condition in eigenmode analysis is incorrectly predicted as having a closed bandgap due to apparent losses associated with radiative modes. Such misinterpretation is particularly critical for Dirac-type unit cells, where accurate identification of bandgap opening or closure is essential. To address these limitations, the use of an appropriate circuit model, together with a robust extraction of scattering parameters that is less sensitive to radiative effects, provides a reliable framework for accurately distinguishing between open and closed bandgap conditions and for precise modeling of the dispersion behavior.

\section{Conclusion}
In this paper, a circuit-based dispersion analysis framework was presented for periodic cross-shaped unit cells with and without a central patch, targeting Dirac-type dispersion characteristics. By systematically comparing eigenmode full-wave results with dispersion diagrams extracted from driven scattering parameters, the limitations of conventional S-parameter-based methods in the presence of radiative modes were highlighted. To address these challenges, an improved equivalent circuit model based on central-patch segmentation and tuned parasitic capacitors was developed, enabling accurate reconstruction of both scattering parameters and dispersion behavior. The close agreement observed between the circuit-based and full-wave results for open- and closed-bandgap conditions confirms the validity and reliability of the proposed approach. Overall, the introduced modeling strategy provides a robust and physically consistent tool for accurate dispersion characterization of periodic structures and offers valuable insight for the design and analysis of Dirac-type unit cells.

\bibliographystyle{IEEEtran}
\bibliography{mybib}

\end{document}